\documentstyle[psfig]{l-aa}

\begin{document}

\thesaurus{03.06.1; Praesepe; 18.02.1; 19.71.1}
\title{Evolution of mass segregation in open clusters: some observational
evidences\thanks{Tables 2 and 3 are available only in electronic form from the
Strasbourg ftp server at 130.79.128.5}}

\author{D.~Raboud \inst{1,2} \and J.-C.~Mermilliod \inst{2}}

\institute{Observatoire de Gen\`{e}ve, CH-1290 Sauverny \and
Institut d'Astronomie de l'Universit\'{e} de Lausanne, CH-1290 
Chavannes-des-Bois}

\date{Received date; accepted date}

\maketitle

\begin{abstract}

On the basis of the best available member list and duplicity information, we 
have studied the radial structure of Praesepe and of the very young open cluster
NGC 6231. We have found mass segregation among the cluster members and between
binaries and single stars, which is explained by the greater average mass of the
multiple systems. However, the degree of mass segregation for stars between 1.5
and 2.3 M$_\odot$ is less pronounced in Praesepe than in the Pleiades. Furthermore,
mass segregation is already present in the very young open cluster NGC 6231
although this cluster is likely still not dynamically relaxed. We discuss the
implications of these results and propose a qualitative scenario for the evolution
of mass segregation in open clusters. 

In Praesepe the mass function of single stars and primaries appears to be
significantly different, like in the Pleiades. We observe an absence of
ellipticity of the outer part of Praesepe. 

\keywords{Clusters: open - Individual: Praesepe; NGC 6231 - Structure - Dynamical
evolution - Star: formation}

\end{abstract}

\section{Introduction}

The dynamical evolution of open clusters has been for a long time mostly a subject
for models and numerical simulations. The currently accepted model
predicts the appearence of mass segregation and concentration of binary stars
towards the cluster centers due to equipartition of kinetic energy among cluster
members, via two-body interactions (see for example the work  of
Spitzer \& Mathieu 1980, Kroupa 1995, de la Fuente Marcos 1996, among others).
As a consequence of this ``standard'' model, one should observe that mass 
segregation increases with cluster age. More precisely, if the cluster is younger
than one relaxation time, it should not exhibit any mass segregation, or only
marginally if the dynamical evolution is producing the segregation.

Observational evidences have so far been difficult to gather because of the
generally too limited cluster surface coverage and the lack of membership criteria.
Only a small number of open clusters are known to present mass segregation. A
study of the radial structure of the Pleiades, using the detailed knowledge of
the membership based on proper motions, radial velocities and photometry
(Raboud \& Mermilliod 1998, referred to as RM98), confirms the presence of an
extended corona around the cluster core and the existence of a clear mass
segregation among cluster members found by van Leeuwen (1983). 
It also shows that the corona boundary is
elliptical and that the mass function for the binary and single stars are
different. Additional open clusters are known to present mass segregation: the
Hyades (Perryman et al. 1997), M11, M35 and M67 (Mathieu 1983, 1984). Results are
similar: the most massive and the multiple stars are always concentrated and the
mass segregation is less important among the low mass stars ($M$  $<$ 1 - 1.5
M$_\odot$) in these clusters.

However very recent studies of extremely young open clusters, such as the Orion
Trapezium (Hillenbrand 1997a) or embedded clusters in star-forming region
(Lada \& Lada 1991) have shown that massive stars are already close to the cluster
center, even at ages of a million years or so. These results, unexpected within
the framework of the ``standard'' mass segregation model, seem to imply that mass
segregation in extremely young objects is unlikely the result of cluster dynamical
relaxation, but may be the result of cluster formation.

To better understand open cluster dynamical evolution it seems important to examine 
the time evolution of mass segregation by covering a large age range. With the
increasing number of observations of clusters at various ages it should be possible
to produce evolutionary scenarii based on observations and provide clues for
comparison with the theoretical predictions, mainly as concerns mass segregation
which is directly observable.
To investigate this issue we have completed extensive observational programs on three
clusters with ages differing by nearly an order of magnitude: NGC 6231, the Pleiades
and Praesepe (M44).

NGC 6231 (3-4 Myr) was analysed by Raboud (1996, 1997) and Raboud et al. (1997).
It has been selected because of its large number of massive early-type stars and
because it seems to be a fully exposed cluster with no embedded parts or links
with residual nebulosity in spite of its young age. It is therefore a very
good example of a very young cluster, probably not much influenced by
stellar or dynamical evolution.

The Pleiades have been analysed in a previous paper (RM98) and a fine mass 
segregation has been clearly established. Although the Pleiades do not
present red giant stars as some clusters of the same age do, their properties
seem representative of the charactristics of similar open clusters, 
e.g. NGC 2422.
However the proximity of the cluster and the large proper motion make it possible
to safely identify cluster members in the outer parts.

Results to be published (Mermilliod \& Mayor 1998) are used to study the structure of 
Praesepe with an approach similar to that used for the investigation of the Pleiades.
The reason to select Praesepe is again its proximity, the large proper
motion and the extensive radial-velocity material obtained over 20 years
of observations which helps defining the membership and identifying the binary
stars.

We first present the available data for Praesepe and NGC 6231 in Sect. 2. We
analyse the structure of the two clusters in Sect. 3. In Sect. 4 we discuss the
comparison between the structures of NGC 6231, the Pleiades and Praesepe.
Section 5 summarizes the main results and concludes the paper.

\section{Observational data}

\subsection{Praesepe}

\subsubsection{Sample}
The complete list of members used in the investigation includes not
only the central part considered by Klein-Wassink (1927, referred to as KW), 
but also a wide surrounding area, out to 4$\degr$ from the center investigated by 
Mermilliod et al. (1990) who identified 43 corona members. The membership
attribution is based on proper motion, radial velocity and photometry.
The discussion of the membership and binarity of about 100  (F5-K0) stars in 
Klein-Wassink (1927) area is in preparation (Mermilliod \& Mayor 1998). The binary 
status has been examined and 25 spectroscopic binaries have been discovered.
Three of them are found in triple systems (Mermilliod et al. 1994). Eighteen
orbits with periods between 3.9 and 7400 days have been determined. 

These two lists form the most complete sample of members earlier than K0 out 
to 4$\degr$ from the cluster center. We cannot claim that all corona members
have been discovered, because the proper motion surveys (Artjukhina 1966a, 1966b) 
are of medium precision. However, the 43 corona members already represent 51\%
of the total number of F5-K0 members. Any new F5-K0 member added to the halo sample
will enhance the observed mass segregation.

\subsubsection{Binarity among the upper main sequence}
The radial-velocity observations for stars on the upper main sequence (MS) are rather
old and partly unpublished. The available information is mostly based on the
radial velocities published by Wilson \& Joy (1950), two or three spectra per star
obtained around 1923.  The individual measurements have been published in Abt's
(1970) compilation of Mt-Wilson observations. Rebeirot (1966) has published mean
values based on the objective prism technique. The results of McDonald (1959) and
Trumpler have not been published, but a copy of the mean values had been kindly
communicated by Hill (1978). Radial velocities (one or two per star) have been
obtained by Dickens et al. (1968), but the Julian dates have not been published.

Table \ref{tab:mem} summarizes the available information. It gives the KW
identification, the results of Wilson \& Joy (1950), mean radial velocities,
standard errors and number of measurements, those of McDonald (1959) and Trumpler,
mean radial velocities and number of spectra, and of Rebeirot (1966), mean radial
velocities, errors and number of plates. The remarks comment on the binary status.
They have been deduced from the notes of each study and from the comparison of the
mean radial velocities obtained at different epochs.

We have also observed several Am stars with the CORAVEL scanner and got an orbit
for three of them, KW 40, 279 and 538. The orbit for the double-lined Am binary
KW 229 (Sanford 1931) has been known for many years. Our observations confirm his
elements. KW 40 is another triple system in Praesepe, with a short period of 6
days, and a long one around 2900 days. The discussion of these orbits will be
presented in another paper.

Multiplicity of the red giants has been discuted by Mermilliod \& Mayor (1989):
KW 428 is a binary with a period of 998 days.

\begin{table}
\caption[]{Literature data for upper MS members in Praesepe.}
\label{tab:mem}
\begin{flushleft}
\begin{tabular}{rrrrrrrrrrrl}
\hline\noalign{\smallskip}
KW &  Wilson & $\sigma$ & n &McD & n & Tr & n & Reb & $\sigma$ & n & Remarks$^{1}$ \\
\hline\noalign{\smallskip}
   38 & 34.1 & 7.2 & 3 &      &   & 32.1 & 3 & 31. & 2.0 & 7 &  \\
   40 & 34.4 & 5.3 & 3 &      &   &      &   &     &     &   & SB1O \\
   45 & 31.0 & 0.6 & 3 & 33.9 & 3 & 32.2 & 4 & 29. & 7.5 & 7 &  \\
   50 & 27.4 & 1.7 & 4 & 34.2 & 4 & 34.7 & 5 & 33. & 7.5 & 4 &  \\
  114 & 34.8 & 4.3 & 3 & 29.0 & 3 & 32.1 & 4 & 36. & 7.5 & 7 &  \\
  124 &      &     &   & 21.7 & 1 & 30.0 & 3 & 20. & 3.5 & 7 & SB1O \\
  143 & 28.4 & 3.8 & 3 & 27.5 & 1 & 32.4 & 4 & 36. & 3.5 & 7 &  \\
  150 & 30.9 &10.2 & 3 & 32.7 & 4 & 30.8 & 3 & 26. & 7.5 & 6 & SB1 \\ 
  203 & 36.5 & 7.4 & 3 & 36.8 & 1 & 29.6 & 4 & 42. & 7.5 & 7 & VB, 0$\farcs$4 \\   
  204 & 30.6 & 3.5 & 4 & 33.4 & 3 & 27.6 & 4 & 37. & 7.5 & 2 &  \\
  207 & 30.7 & 9.7 & 3 & 35.3 & 2 & 34.6 & 4 & 24. & 7.5 & 7 & SB1 \\
  218 &      &     &   &      &   & 27.6 & 4 & 35. & 3.5 & 7 &  \\
  226 &      &     &   &      &   & 39.0 & 5 & 31. & 2.0 & 7 &  \\
  229 &      &     &   & 36.3 & 2 & 31.7 & 3 & 26. & 3.5 & 5 & SB2O \\
  265 & 33.1 & 3.0 & 3 & 35.4 & 2 & 32.6 & 3 &     &     &   & OccB, 0$\farcs$4 \\
  271 &      &     &   &      &   & 38.6 & 5 & 37. & 2.0 & 2 & PHB \\
  276 & 38.5 & 4.8 & 4 & 27.1 & 3 & 31.1 & 4 & 39. & 2.0 & 6 & SB1 \\
  279 & 16.3 &14.1 & 3 & 19.2 & 2 &      &   & 22. & 2.0 & 2 & SB1O \\
  284 & 30.7 & 2.9 & 4 & 28.7 & 3 & 39.9 & 5 & 42. &     & 1 & OccB, 0$\farcs$11 \\
  286 & 27.3 & 1.0 & 3 &      &   & 38.4 & 5 & 28. & 3.5 & 7 &  \\
  292 &-22.0 &94.6 & 2 &      &   &      &   & 26. & 3.5 & 7 & SB2 \\
  295 &      &     &   &      &   & 34.4 & 3 &     &     &   &  \\
  300 & 20.7 &27.3 & 8 & 59.3 & 1 & 38.2 & 2 &     &     &   & SB2 \\
  318 &      &     &   &      &   & 39.7 & 3 &  5. & 3.5 & 7 &  \\
  323 & 37.1 & 5.9 & 3 & 34.8 & 1 & 32.8 & 3 & 34. & 2.0 & 7 & PHB \\ 
  328 & 36.8 & 2.8 & 3 & 21.4 & 1 & 35.5 & 3 & 2.  &     & 1 & SB2 \\
  340 &      &     &   &      &   & 31.7 & 4 & 31. & 2.0 & 7 &  \\ 
  348 & 27.6 & 5.5 & 6 & 38.5 & 3 & 30.3 & 4 & 19. &     & 1 &  \\
  350 &      &     &   &      &   & 33.0 & 3 & 29. & 3.5 & 7 &  \\
  370 &      &     &   &      &   & 32.5 & 3 & 33. & 3.5 & 7 &  \\
  375 & 33.2 & 8.2 & 2 &      &   & 29.5 & 5 & 34. & 3.5 & 7 & SB1 \\
  385 & 40.8 &20.8 & 3 &      &   &      &   & 34. & 3.5 & 7 & VB, 2$\farcs$1 \\ 
  411 &      &     &   &      &   &      &   & 29. & 7.5 & 7 &  \\
  429 &      &     &   &      &   &      &   & 29. & 7.5 & 7 &  \\
  445 & 27.1 & 1.3 & 2 & 36.1 & 3 & 36.8 & 3 & 33. & 7.5 & 7 &  \\
  449 &      &     &   & 24.0 & 2 & 36.6 & 3 &     &     &   &  \\
  459 &      &     &   &      &   & 31.2 & 4 & 23. & 3.5 & 7 &  \\
\noalign{\smallskip}
\hline
\multicolumn{12}{l}{$^{1}$ spectroscopic binary: SB1, SB2, with orbit: SBO;} \\
\multicolumn{12}{l}{\,\, photometric binary: PHB; visual binary: VB;} \\ 
\multicolumn{12}{l}{\,\, occultation binary: OccB} \\
\end{tabular}
\end{flushleft}
\end{table} 

\subsubsection{The member catalogue}
Table 2 summarizes the available data for our working list of 185 Praesepe members.
It contains the star identification, $V$ and $B-V$ from BDA, the open cluster
database (Mermilliod 1995), the $x$ and $y$ rectangular positions in arc minutes,
the distance from the cluster center, the multiplicity status, remarks and the
deduced masses of the stars and components. Table 2 is available only in
electronic form from the Strasbourg anonymous ftp server (130.79.128.5).
For single stars the individual masses have been determined from the B-V colour
index with a power law relation between mass and B-V derived from an isochrone
computed from the models of Schaller et al. (1992) for log t = 8.92 and z = 0.02.
For the binary stars presenting a marked displacement above the ZAMS in
the colour-magnitude diagram, we have proceeded as explained in a similar
discussion of the Pleiades (RM98).

\subsection{NGC 6231}

\subsubsection{Sample}

NGC 6231 is one of the richest and youngest exposed open cluster known
(Mermilliod 1981; Meynet et al. 1993). It is located at $l$=343\fdg 5 and
$b$=1\fdg 2. This cluster is found near the southern end of the very young
association Sco OB1 and is usually considered as its nucleus (Perry et al.
1991).

However, the amount of data concerning NGC 6231 is beyond comparison with
those available for Praesepe or the Pleiades, mainly because of its larger 
distance (1.8 kpc, Raboud et al. 1997), although efforts have been done to improve the
data. Raboud (1996) investigated the binarity among B-type stars from ESO 3.6m radial
velocities. He derived a minimum binary fraction of 52 \% in the considered
population. Raboud et al. (1997), using Geneva photometry, identified 64 new members
out to a distance of 13\arcmin\, from the center, extending the Seggewiss area
(8\arcmin). References to previous works on NGC 6231 can be found in these two papers.

As a consequence of the small amount of available data, we concentrate on the
existence of mass segregation in this very young open cluster.

\subsubsection{The member catalogue}

We have composed a large table collecting all 300 members brighter than
$V_{0} = 12.5$ from Raboud et al.'s (1997) catalogue, completed with $UBV$
(Seggewiss 1968, Garrison \& Schild 1979) and $uvby$ (Balona \& Laney 1995)
photometric data. 192 stars are measured in the Geneva system, 66 in $UBV$
and 42 in $uvby$. Table 3 contains the star identification, $V$ and $(B-V)$,
$[B-V]$ or $b-y$ depending on the photometry used, the photometric system
used, the $x$ and $y$ rectangular positions in arc minutes, the distance from
the cluster center, the multiplicity status, remarks and the deduced masses of
the stars and components. Table 3 is available only in electronic form from
the Strasbourg anonymous ftp server (130.79.128.5).

The individual masses of the stars have been derived by different techniques
depending on the photometry used and on the multiplicity status. For stars
measured in the Geneva system, we derived the temperature with the calibration
of the $X$ parameter (Cramer 1984). The mass is then obtained with an isochrone
calculated by the models of Schaller et al. (1992) for log t = 6.6 and z = 0.02.
This method is only valid for the hotter stars. However, for temperatures
greater than log $T_{\mbox{eff}}$ = 4.5 the relation between the mass and the
temperature becomes too vertical and the mass determination failed. In these
cases we only considered a lower mass limit of 22 M$_\odot$ for the stars.

For stars outside the calibration range of the $X$ parameter, we used the
relation bewteen $[B-V]_{0}$ and the mass derived from the same isochrone. A
similar technique is used for stars measured in $UBV$. For the $uvby$ data we
compute the temperature from $(b-y)_{0}$, with the calibration of Hauck \&
K\"{u}nzli (1996). The mass is derived with the isochrone defined above.

All these techniques are valid \textit{only} if the stars are on the main sequence.
Raboud et al. (1997) showed the existence of a candidate pre-main
sequence (PMS) population. The mass of the stars belonging to this population
have been estimated with the PMS evolutive tracks from Bernasconi (1996).

The derivation of the multiple star components make use of similar techniques
as the ones described in RM98.

\section{Structure}

\subsection{Praesepe}

This section presents a study in every respect similar to that described in RM98
for the Pleiades, to facilitate comparisons. Therefore we give
here only the results with a minimum of details. The reader is referred to RM98
for the explanations of the various methods and formulae used, and for precise
definitions.

\subsubsection{Global overview}

\begin{figure}[t]
\centerline{\psfig{figure=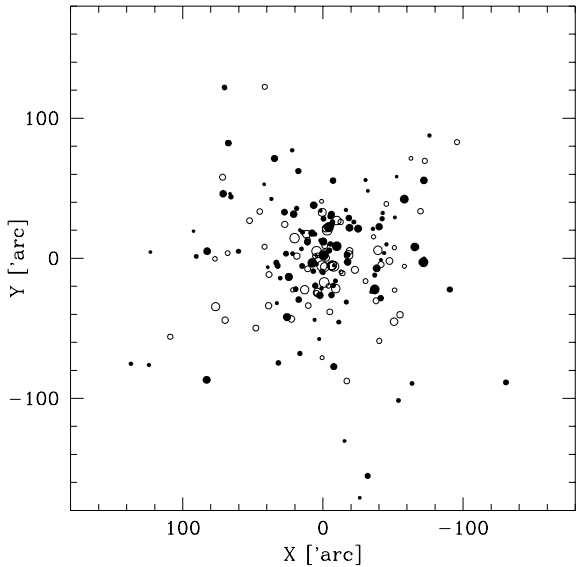,height=80mm,width=80mm}}
\caption[]{Map of Praesepe displaying all the member stars considered in
this study. Open circles are multiple stars and filled circles stand for single
stars. The point sizes are related to the magnitudes of the stars or systems.
North is at the top and East at the left of the map.}
\label{fig:cartemasse}
\end{figure}

From the data collected in Table 2, we have computed the cluster mass center:
$\alpha_{1950}=8^{h}37^{m}32^{s}$; $\delta_{1950}$=19\degr48\farcm8\, and used it
to plot a chart of the Praesepe cluster (Fig. \ref{fig:cartemasse}) displaying the
single stars as filled circles and the multiple ones as open circles.

The cluster appears circular out to 3\degr\, from its center. In order to compare
the overall shape of Praesepe with that of the Pleiades (RM98) we divided the two
clusters in several sectors and computed an \textit{asymmetry estimator} (Bouvier
1961) defined as
\begin{equation}
D=\frac{1}{N}\sum_{i=1}^{p}{\left|n_{i}-\overline{n}\right|},
\end{equation}
where $N$ is the total star number, $p$ is the number of sectors, $n_{i}$ is
the number of stars in the $i$th sector and $\overline{n}=N/p$. $D$ varies in
the interval [0,2[, with $D$=0 corresponding to a circular cluster and
$D \rightarrow$ 2 corresponding to a linear arrangement of the ``cluster'' stars.

Table \ref{table:D} displays the asymmetry-estimator values for two cluster
subdivisions (8 and 12 sectors, respectively). In both cases, Praesepe appears
clearly more circular than the Pleiades. Therefore the known halo of Praesepe
do not present any ellipticity such as that observed for the Pleiades (RM98). 

\setcounter{table}{3}
\begin{table}
\caption{Comparison between the asymmetry estimators for Praesepe and the
Pleiades.}
\label{table xx}
\begin{flushleft}
\begin{center}
\begin{tabular}{|l c c|}
\hline \rule{0pt}{1.2em}
{Cluster} & {$p$=8} & {$p$=12}\\
\hline
\textbf{Praesepe}   & 0.124 & 0.135 \\
\textbf{Pleiades}   & 0.215 & 0.207 \\
\hline
\end{tabular}
\end{center}
\end{flushleft}
\label{table:D}
\end{table}

Following Wielen (1975), the ratios of the three orthogonal axes of a cluster,
considered  as a tridimensional ellipsoid, should be 2.0:1.4:1.0. The larger axis
is pointing towards the galactic center and the smaller one is perpendicular to
the galactic disk. As Praesepe lies at a galactic longitude of 205$\fdg$5 and is
close to us (158 pc), we only observe the ratio of the second and third axes,
namely 1.4:1.0, which corresponds to an ellipticity of 0.29. However,  Praesepe
is $\sim$ 85 pc above the galactic plane and we see it under an angle of 32$\fdg$5.
We thus observe an effective axial ratio of 1.02:1.0, which corresponds to an
ellipticity of only 0.02. This theoretical expectation agrees with our observation of
a projected round-shaped halo for Praesepe.

However, the absence of ellipticity of the outer part of Praesepe could also be
real and not an artifact of projection effects. If real, the round-shaped halo
may be the signature of multiple interactions between the cluster and interstellar
clouds, because such gravitational interactions rapidly stripped off the outermost
halo stars, which filled the elliptical region allowed by the galactic tidal field.
This scenario (Wielen 1974) predicts that the oldest clusters should have the
more circular halos, because they have statistically suffered more encounters
with interstellar clouds than younger clusters.

Nowadays, Praesepe is the oldest open cluster for which we have a characterisation
of its halo shape, which is consistent with the theoretical expected ellipticity.
The Pleiades (RM98), NGC 3532 (Gieseking 1981) are younger, the Hyades (Oort 1979)
have the same age, and they all present elliptical outerparts. Therefore it would
be very interesting to investigate whether clusters older than Praesepe have
elliptical halos or not. Such investigations would constrain the frequency of
gravitational encounters between open clusters and interstellar clouds.

\subsubsection{Mass segregation}

\begin{figure}[thb]
\centerline{\psfig{figure=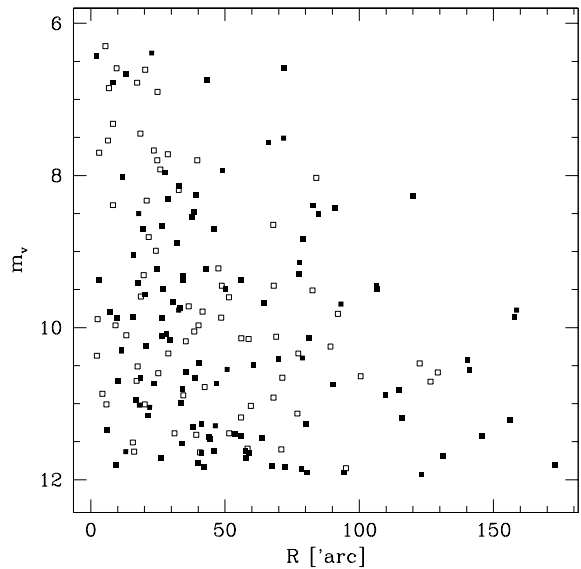,height=80mm,width=80mm}}
\caption[]{Apparent magnitudes of the stars as a function of their radial distances
to Praesepe center. The open and filled squares denote multiple and single stars
respectively.}
\label{fig:VvsR}
\end{figure}

\begin{figure}[thb]
\centerline{\psfig{figure=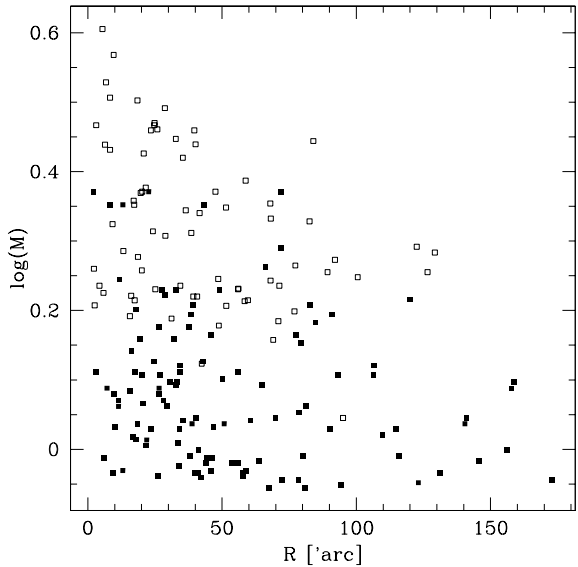,height=80mm,width=80mm}}
\caption[]{Logarithm of the star masses as a function of their radial distances to
Praesepe center. This diagram is a direct representation of mass segregation. Same
symbols as in Fig. \ref{fig:VvsR}.}
\label{fig:lMvsR}
\end{figure}

The concentration of multiple stars relative to single ones, of bright stars
relative to fainter ones, and of massive stars relative to less massive ones is
apparent in Figs \ref{fig:VvsR} and \ref{fig:lMvsR}. 
These two figures represent the radial extension of stars of different
magnitudes or masses and also show the completeness status of our survey in terms of 
magnitude, mass and radial extensions. Fig.~\ref{fig:lMvsR} very clearly demonstrates
that the size of the cluster increases when the mean stellar mass decreases. The
trend is suprisingly rather well defined.
As a consequence the definition of cluster radii is not simple and visual estimates
of this parameter on photographs is very sujective and prone to large errors
depending on the density of the stellar background on which the cluster is projected.
There is only one star more massive than 2 M$_\odot$ out of the limit defined by 
the trend (lying out to 2\degr\, from the center). Although in principle
stars of any mass could be found anywhere inside the cluster boundaries, it appears
that energy equipartition confines the stars in bounded volumes. Are the stars 
observed outside the normal boudaries being ejected and leaving the cluster,
although they are still located within the tidal radius ? 

To investigate more accurately the radial distribution of the different star
populations we have split the sample into classes, according to the stellar
multiplicities and to the stellar masses.

\paragraph{Single and multiple stars:}

\begin{figure}[thb]
\centerline{\psfig{figure=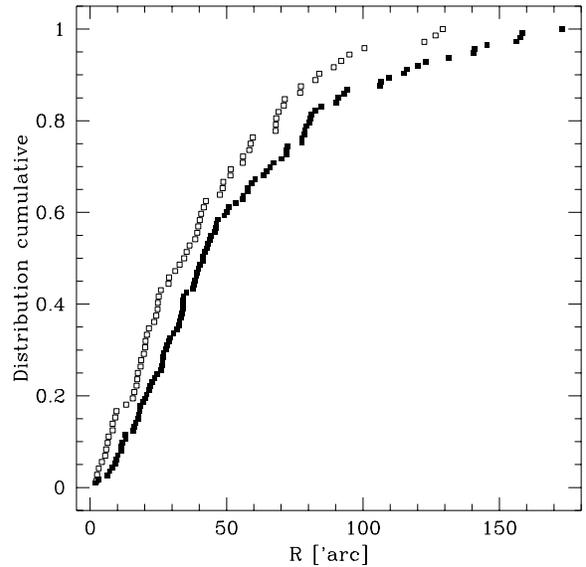,height=80mm,width=80mm}}
\caption[]{Cumulative distributions for the multiple stars (open squares) and the
single stars (filled squares) in Praesepe.}
\label{fig:cumsmplsb}
\end{figure}

The multiple systems are clearly more concentrated towards the cluster center
than the single stars (Fig. \ref{fig:cumsmplsb}). A Kolmogorov-Smirnov test
indicates that the probability of false rejection of the null hypothesis, i.e.
that the two distributions are identicals, is 9.4 \%.

\begin{figure}[thb]
\centerline{\psfig{figure=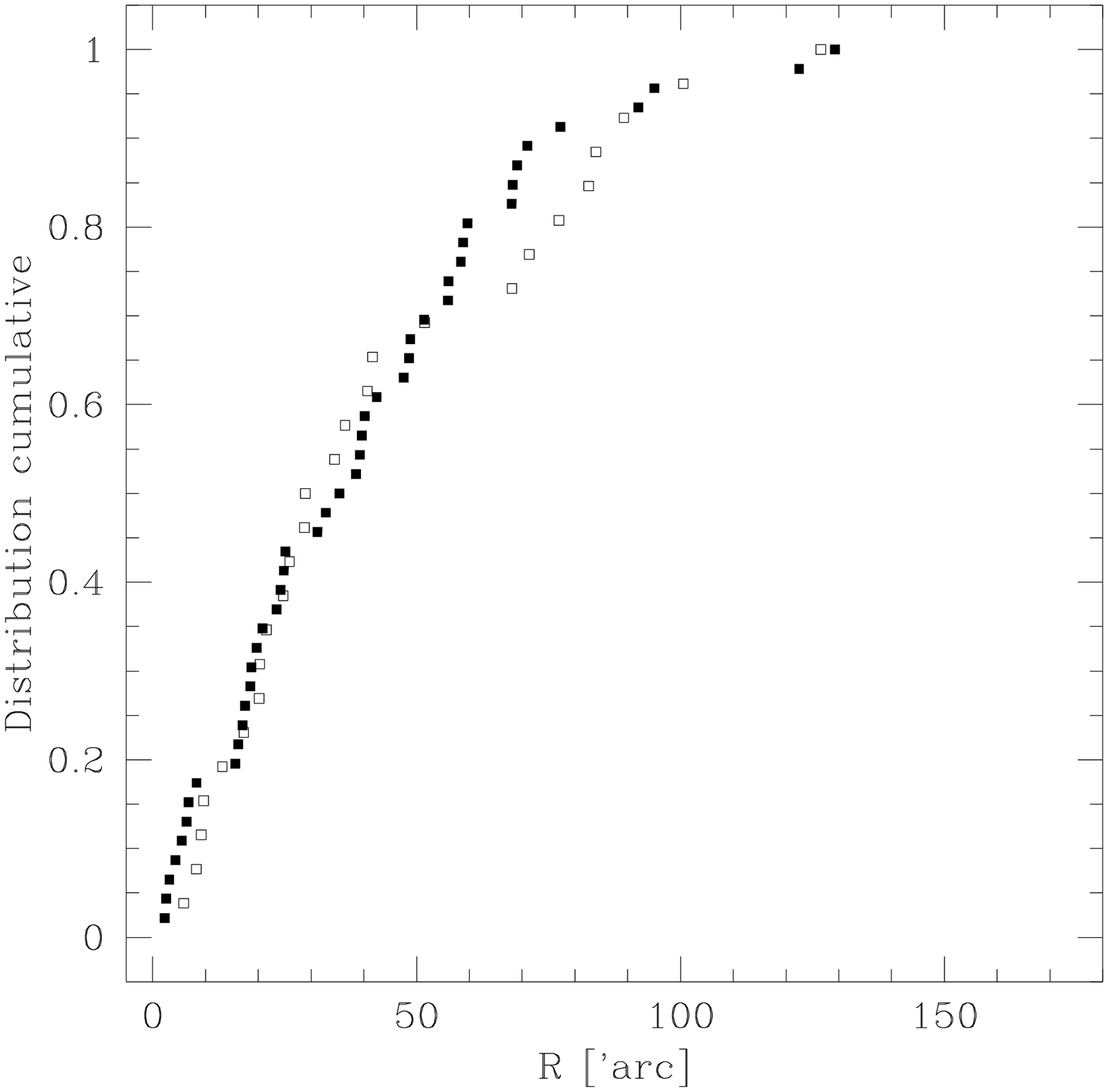,height=80mm,width=80mm}}
\caption[]{Cumulative distributions for the ``short period'' binaries (filled
squares) and the ``long period'' binaries (open squares) in Praesepe. See text for
more details.}
\label{fig:cumsbvb}
\end{figure}

Among the multiple star population itself we can divide the sample between
``short period'' binaries (spectroscopic) and ``long period'' binaries (visual,
occultation and photometric binaries). The resulting two cumulative distributions
are plotted in Fig. \ref{fig:cumsbvb}. Their radial distributions are very similar.

These results agree with the hypothesis that the radial segregation of binaries
towards the cluster center depends mainly on the total mass of the systems, and
not on their periods (at least for periods smaller than $\sim$ 10$^3$ yr). The same
conclusion was obtained and discussed by Raboud \& Mermilliod (1994) and by RM98
in the Pleiades.

\paragraph{Sample subdivision using mass criteria:}

\begin{figure}[thb]
\centerline{\psfig{figure=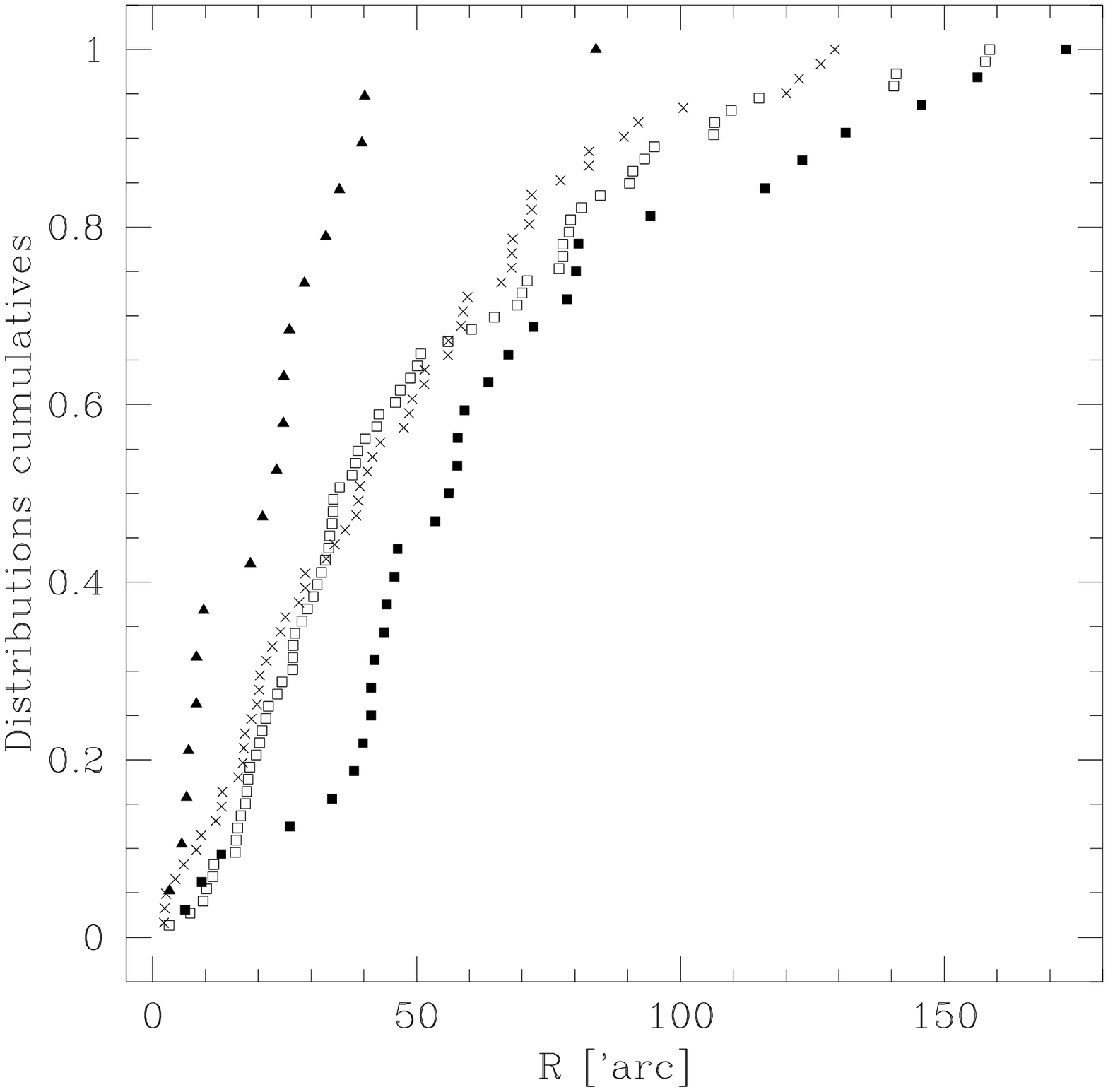,height=80mm,width=80mm}}
\caption[]{Cumulative distribution functions for different mass intervals in
Praesepe: $M < 1$ M$_\odot$ (filled squares), $1 < $M$ < 1.6$ M$_\odot$ (open
squares), $1.6 < $M$ < 2.5$ M$_\odot$ (crosses) and $M > 2.5$ M$_\odot$ (filled
triangles).}
\label{fig:cumul4a}
\end{figure}

To characterize the degree of mass segregation among different populations in the
cluster, we subdivise the sample into 4 groups. Fig. \ref{fig:cumul4a} represents
the cumulative distributions for 4 mass intervals, chosen identical as those used
in the study of the Pleiades (RM98).

The most massive stars in Praesepe are obviously more concentrated than stars of
smaller mass, while the stars with masses less than 1.0 M$_\odot$ are the least
concentrated. The important and new point is the similarity between the radial
distributions of the two intermediate mass intervals. In this mass interval
(1.0 $<$ $M$ $<$ 2.5 M$_\odot$) we find the surprising result that the degree
of mass segregation is less pronounced in Praesepe than in the Pleiades, in spite
of the greater age of the former cluster.

\subsubsection{Characteristic radii}

We derived the values of the radius containing half of the total number of stars
($r_{n/2}$), the radius containing half of the total mass of the stars ($r_{m/2}$),
the core ($r_{c}$), the tidal ($r_{t}$) and the harmonic radii ($\overline{r}$).
These derivations were done for different member sub-samples, following the
procedures described in RM98. The results are presented in Table \ref{tab:radii},
with their uncertainties indicated in brackets.

\begin{table}
\caption{Characteristic radii [\arcmin] for different member sub-samples in
Praesepe. The errors associated with $r_{n/2}$ and $r_{m/2}$ are typically
between 3 and 7 [\arcmin].}
\label{tab:radii}
\begin{flushleft}
\begin{center}
\begin{tabular}{|l c c c c r|}
\hline \rule{0pt}{1.2em}
{Population} & {$r_{n/2}$} & {$r_{m/2}$} & {$r_{c}$} & 
\multicolumn{1}{c}{$r_{t}$} & \multicolumn{1}{c|}{$\overline{r}$} \\
\hline
\textbf{Complete sample}   & 39 & 34 & 22 (10) & 242 (107) &  65 (13) \\
Bright stars               & 28 & 25 & 16 (14) & 227 (346) &  47 (19) \\
Faint stars                & 42 & 41 & 25 (16) & 271 (205) &  73 (20) \\
Massive stars              & 33 & 29 & 17 (13) & 248 (312) &  53 (17) \\
Less massive stars         & 43 & 42 & 27 (18) & 270 (224) &  74 (22) \\
Single stars               & 41 & 40 & 24 (15) & 271 (199) &  71 (19) \\
Multiple stars             & 34 & 28 & 19 (17) & 238 (350) &  53 (19) \\
\hline
\end{tabular}
\end{center}
\end{flushleft}
\end{table}

We adopt cut-off values similar to those used for the Pleiades (RM98), namely
$V$=9.6 and $M$=1.5 M$_\odot$, to subdivise the bright/faint and the massive/less
massives stars respectively.

The various parameters describing Praesepe are (for a distance of 158 pc):
$r_{c}$ = 1 $\pm$ 0.5 pc, $r_{t}$ = 11.1 $\pm$ 5 pc, $r_{n/2}$ = 1.8 $\pm$
0.2 pc, $r_{m/2}$ = 1.6 $\pm$ 0.2 pc and $\overline{r}$ = 3 $\pm$ 0.6 pc.

\subsubsection{The mass functions}

Due to the improved material collected in this paper, we are able to compute
various mass functions for Praesepe. This will allow us to test the effect of
unresolved binaries, of the radial extension of the survey and to confirm whether
the mass function for single stars is different from the mass function of the
primaries of multiple systems, as it was observed in the Pleiades (RM98).

We fitted a Salpeter-type power law in the form
\begin{equation}
log\left(\frac{df}{dM}\right)=C-(1+x)\;log(M)
\label{eq:fctm}
\end{equation}
throughout the observed data. In Eq.~(\ref{eq:fctm}) $df/dM$ is the number of
stars per unit mass as a function of mass $M$. $C$ is a constant and $(1+x)$ is
the power law exponent, which has the value of 2.35 following Salpeter (1955).

The slope derived without any correction for the binary content has a value of
2.3 $\pm$ 0.4 (case 1 in Table \ref{tab:slopes}), which agrees with the Salpeter
one.

\begin{table}
\caption{Values of different power law exponents $(1+x)$ in Praesepe. The cluster
inner part is the central 2-pc disk, and the cluster outer part is outside this
disk.}
\label{tab:slopes}
\begin{flushleft}
\begin{center}
\begin{tabular}{|l c|}
\hline \rule{0pt}{1.2em}
{Sample} & {$(1+x)$} \\
\hline
(1) Complete sample (with unresolved binaries)           & 2.3 $\pm$ 0.4 \\
(2) Complete sample (singles + primaries)       & 2.5 $\pm$ 0.3 \\
(3) Singles                   & 2.8 $\pm$ 0.3 \\
(4) Primaries                 & 2.1 $\pm$ 0.3 \\
(5) Cluster inner part (singles + primaries) & 1.6 $\pm$ 0.4 \\
(6) Cluster outer part (singles + primaries) & 3.6 $\pm$ 0.7 \\
(7) Complete sample (mass summed)           & 1.2 $\pm$ 0.2 \\
(8) Cluster inner part (mass summed)        & 0.6 $\pm$ 0.1 \\
\hline
\end{tabular}
\end{center}
\end{flushleft}
\label{x}
\end{table}

The slope derived for the single stars and the primaries of multiples systems,
using the available information about them, is 2.5 $\pm$ 0.3 (case 2 in Table
\ref{tab:slopes}). This value is in agreement with the previous one and confirms
the result obtained in the Pleiades (RM98): the determination of the mass-function
slope is not seriously affected by unresolved multiple stars.

\begin{figure}[thb]
\centerline{\psfig{figure=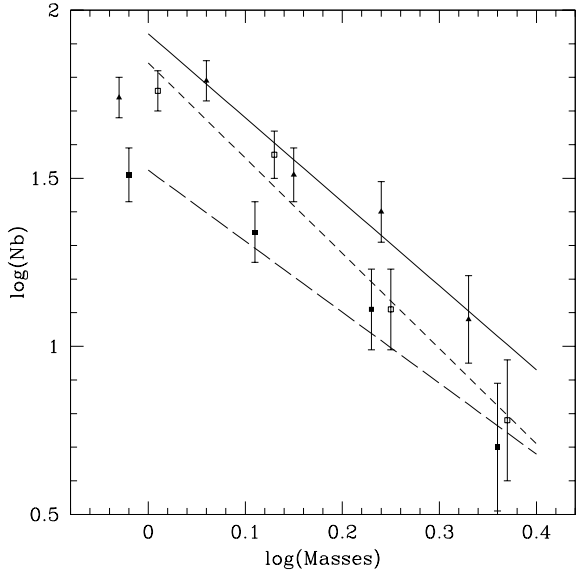,height=80mm,width=80mm}}
\caption[]{Mass functions in Praesepe. The solid line stands for the complete
sample (single stars and primaries). The long-dashed line represents the mass
function of the primaries and the short-dashed line stands for the single stars.}
\label{fig:fmas}
\end{figure}

We are also able to compare the mass function of single stars (case 3 in Table
\ref{tab:slopes}) and of primaries of multiple systems (case 4 in Table
\ref{tab:slopes}). We observe that the slope of the single star mass function
is steeper than that for the primaries (Fig. \ref{fig:fmas}). This result implies
that the components of binary systems are not drawn independantly from the same
mass function as that of single stars. This was already reported for the Pleiades
(RM98) and was qualitatively explained by dynamical evolutionary effects.
Praesepe, older than the Pleiades, has probably undergone a more complete
dynamical evolution and the encounters between single stars and binaries,
leading to the capture of the more massive stars into the multiple systems
(Mathieu 1985), have had time to flatten the primary mass function.

\subsubsection{The frequency of multiples star systems}

Our direct detections of multiple systems give a proportion of 43\% of binaries
in the central 2-pc disk and of 34\% in the outer part of the cluster.
The difference between the inner and outer parts is less pronounced than that
observed for the Pleiades (RM98), i.e. 48\% in the central 2-pc disk, 20\%
outside, although Praesepe overall binary fraction (39\%) is larger than that of
the Pleiades (32\%).

It is interesting to compare these results with the work of Kroupa \& Tout (1992).
Their analysis of the photometric colour-magnitude diagram in Praesepe yielded a
large binary frequency. Values close to unity would be still acceptable although
smaller fractions could not be excluded. 

\subsubsection{Estimation of the cluster total mass}

We have shown (RM98) that the estimation of the cluster total mass is a very
difficult task producing results with large uncertainties. We derive values
between 157 and 3970 M$_\odot$ for the total mass of Praesepe (Table
\ref{tab:masse}). These results were obtained using either the integration of
the cluster mass function, including the contribution of binary companions, or
the relation between the tidal radius and the cluster mass (King 1962):
\begin{equation}
M_{c}=\frac{4A(A-B)}{G}r_{t}^{3}
\label{eq:mc}
\end{equation}
where $G$ is the gravitational constant, $r_{t}$ is the tidal radius of the
cluster, $A$ and $B$ are Oort's constants of galactic rotation. 
The tidal radius considered in Eq.~(\ref{eq:mc}) is measured in the direction
of the galactic center. However, we only observe the tidal radius perpendicular
to this direction but parallel to the galactic disk. In Sect. 3.1.1 we found that
Praesepe may have a flattening close to the theoretical expected one. We could
therefore consider that we have a cluster with axes ratios 2:1.4:1 and then the
tidal radius in the direction of the galactic center has a value of $2/1.4$ times
the value of the observed tidal radius (242\arcmin\, from Table \ref{tab:radii}).
Finally, if we sum up all the stellar masses derived for our whole sample of
stars, which is the only way to properly compute the cluster mass, we obtain
300 M$_\odot$.

\begin{table}
\caption{Results of the different Praesepe total mass determinations (see
text for more details).}
\begin{flushleft}
\begin{center}
\begin{tabular}{|l c c|}
\hline \rule{0pt}{1.2em}
{Method} & \multicolumn{1}{c}{Cluster total mass} & \multicolumn{1}{c|}{1
$\sigma$ confidence interval}\\
 & {M$_\odot$} & {M$_\odot$}\\
\hline
\textbf{Tidal radius}   & 1330 & [229, 3970] \\
\multicolumn{3}{|l|}{(with correction for the cluster flatness)} \\
\textbf{Tidal radius}   & 440 & [\, 157, 987] \\
\multicolumn{3}{|l|}{(without correction for the cluster flatness)} \\
\textbf{Mass function} & \, 590 & [\, 416, 900]\\
\textbf{Summed mass} & \, 300 & \\
\hline
\end{tabular}
\end{center}
\end{flushleft}
\label{tab:masse}
\end{table}

\subsection{NGC 6231}

\subsubsection{Completeness of the member list}

\begin{figure}[thb]
\centerline{\psfig{figure=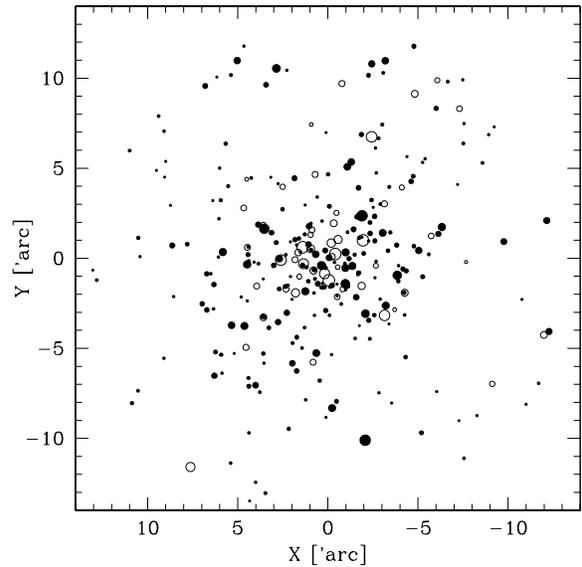,height=80mm,width=80mm}}
\caption[]{Map of NGC 6231 displaying all the member stars considered in
this study. Open circles are multiple stars and filled circles stand for single
stars. The point sizes are related to the magnitudes of the stars or systems.
North is at the top and East at the left of the map.}
\label{fig:carte6231}
\end{figure}

From the data collected in Table 3, we have computed the cluster mass center and
we considered star S320 (Seggewiss 1968) as lying at the center: $\alpha_{1950}=
16^{h}50^{m}41^{s}$; $\delta_{1950}$=-41\degr44.9. Figure \ref{fig:carte6231} shows
a chart of NGC 6231 displaying all the member stars considered
in the present study. The filled and open circles represent single and multiple
stars respectively. The size of the circles is proportional to the star magnitude.

\begin{figure}[h]
\centerline{\psfig{figure=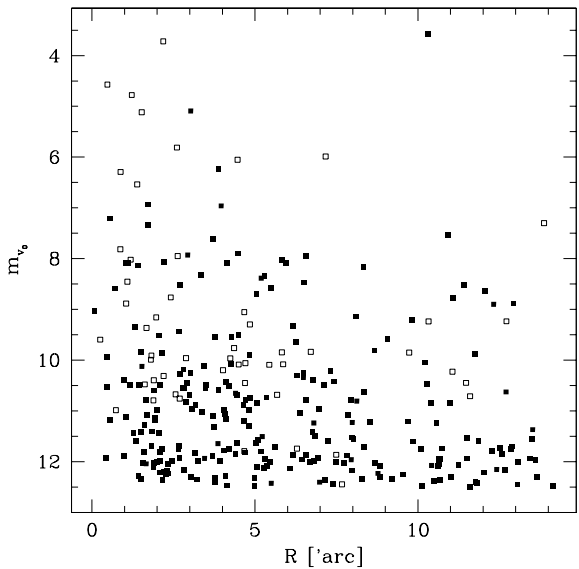,height=80mm,width=80mm}}
\caption[]{Apparent magnitudes of the stars as a function of their radial
distances to the center of NGC 6231. The open and filled squares denote
multiple and single stars respectively.}
\label{fig:VvsR6231}
\end{figure}

Figure \ref{fig:VvsR6231} displays the increase of the cluster size with the
decrease of the star luminosity, which is equivalent to a decrease of mean 
stellar masses.
Although the exact radius at any mass is somewhat fuzzy, probably because 
of the presence in the sample of a few stars belonging to the association
surrounding NGC 6231, mass segregation is clearly present in this diagram.
The massive stars ($V <$ 7) occupy a much smaller volume than the less
massive stars ($V >$ 10).
It will be more quantitatively characterized in the next section.

The status of the outlying stars, i.e. those stars found at larger distance
from the cluster center than the bulk of stars at the same magnitude is worth
considering. In particular the membership of the nine bright stars (Nos 501, 723,
724 726, 745, 749 769, 774 and 810) of the cluster corona (i.e. the region between
$\sim$8.5 and $\sim$13.5 ['arc]) has been discussed in Raboud (1997). These
stars, considered as cluster members from the photometric analysis, are probably
member of the Sco OB1 association. An estimate of the stellar contamination due
to the association was done using the Guide Star Catalogue (GSC), for stars
brighter than $V$=11.4 (corresponding to the fainter reddened magnitude of the
9 discussed stars). It results that the bright stars of the cluster corona have
a probability of more than 50\% of belonging to the association rather than to
NGC 6231. In the following discussion we shall consider both cases, with and
without these stars.

\subsubsection{Mass segregation}

\paragraph{Single and multiple stars:}

\begin{figure}[thb]
\centerline{\psfig{figure=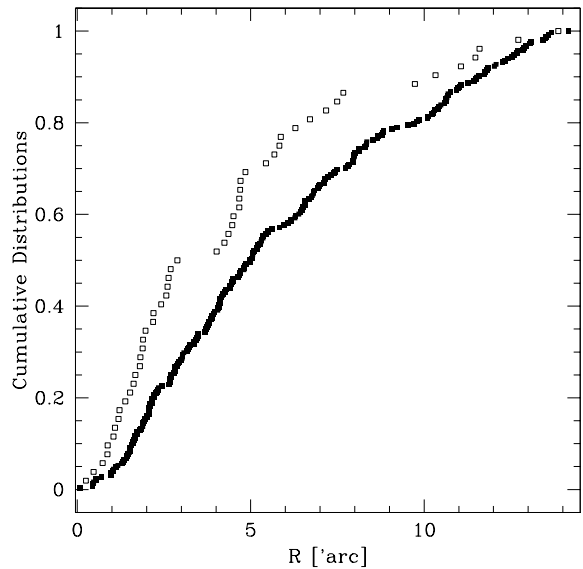,height=80mm,width=80mm}}
\caption[]{Cumulative distributions for all the multiple stars (open squares)
and the single stars (filled squares) in NGC 6231.}
\label{fig:segregationsbsmpl6231}
\end{figure}

\begin{figure}[thb]
\centerline{\psfig{figure=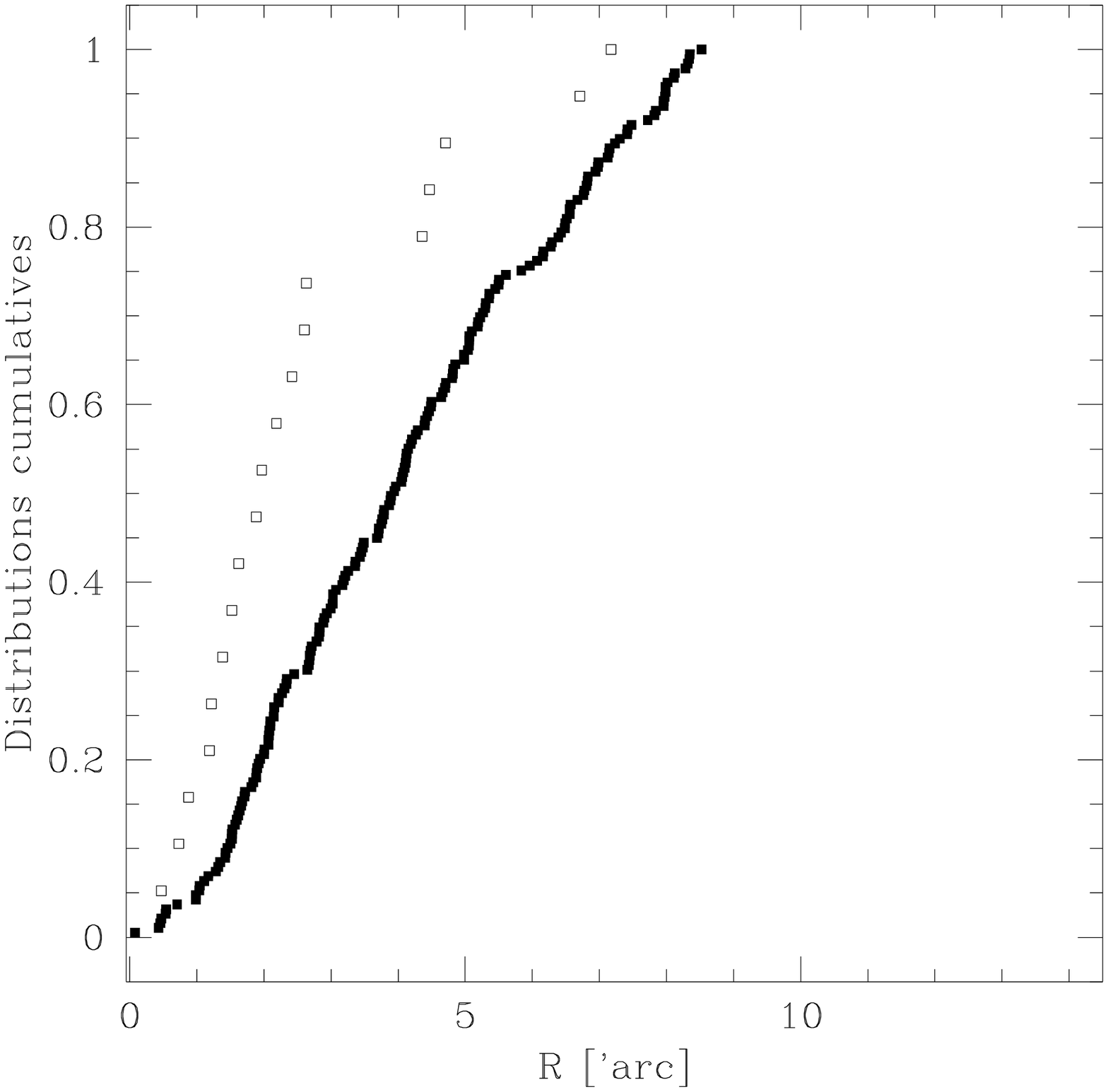,height=80mm,width=80mm}}
\caption[]{Cumulative distributions for the multiple stars, spectroscopically
detected, (open squares) and the single stars (filled squares) in NGC 6231.}
\label{fig:segregationsbsmpl6231b}
\end{figure}

Figure \ref{fig:segregationsbsmpl6231} shows the cumulative distributions for
the multiple stars (open squares) and the single stars (filled squares). The
multiplicity status was derived from radial velocity studies (Levato \& Morrell
1983; Perry et al. 1990; Raboud 1996) and photometric criteria. Because of the
low efficiency of photometric criteria on the steep upper main sequence, Fig.
\ref{fig:segregationsbsmpl6231b} present the distributions of only spectroscopic
multiple stars and single ones in the cluster central 8$\arcmin$, where
spectroscopic information is available. NGC 6231 presents the unique feature that
8 among the 10 brighter stars are spectroscopic binaries with periods shorter
than 6 days (Hill et al. 1974; Levato \& Morrell 1983)

In both figures (\ref{fig:segregationsbsmpl6231} and \ref{fig:segregationsbsmpl6231b})
the multiple stars appear to be more concentrated
than the single ones. Kolmogorov-Smirnov tests clearly confirm that the two
distributions are differents. For Fig. \ref{fig:segregationsbsmpl6231} and
\ref{fig:segregationsbsmpl6231b} respectively: the probabilities are of 0.2 \% and
1.2 \% to reject the null hypothesis, that the two distributions are the same,
even though it is true.

\paragraph{Sample subdivision using mass criteria:}

\begin{figure*}[thb]
\centerline{\psfig{figure=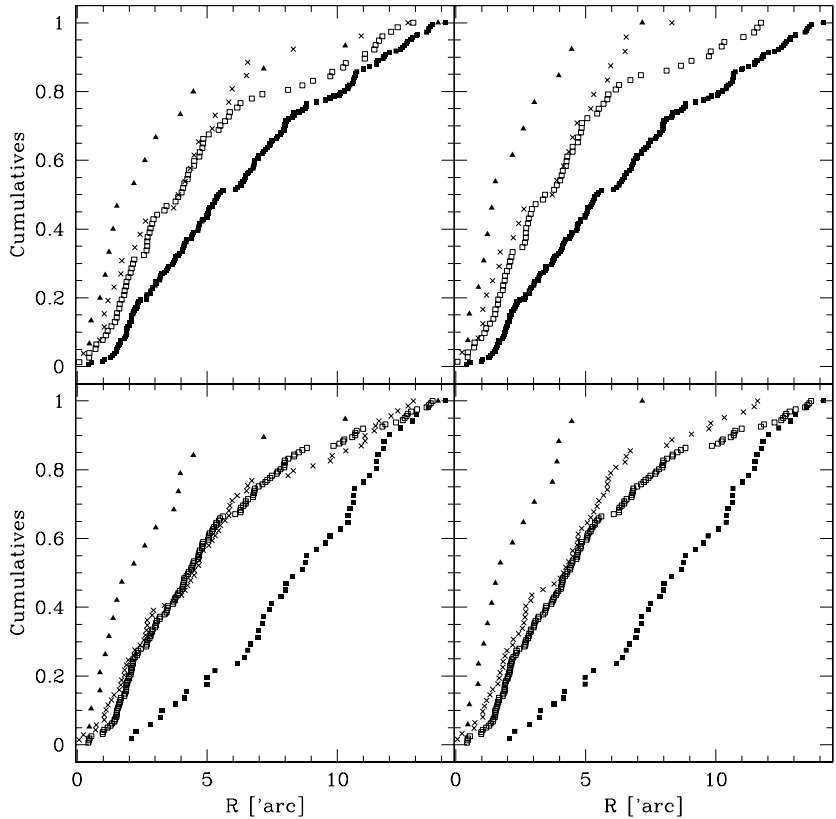,height=120mm,width=120mm}}
\caption[]{Cumulative distributions for two mass interval sets in NGC 6231. For
the two top figures: $M$ $<$ 5 M$_\odot$ (filled squares); 5 $\leq$ $M$ $<$ 10
M$_\odot$ (open squares); 10 $\leq$ $M$ $<$ 20 M$_\odot$ (crosses) et $M$ $\geq$
20 M$_\odot$ (triangles). For the two bottom figures: $M$ $<$ 2.5 M$_\odot$
(filled squares); 2.5 $\leq$ $M$ $<$ 6.3 M$_\odot$ (open squares); 6.3 $\leq$
$M$ $<$ 15.8 M$_\odot$ (crosses) et $M$ $\geq$ 15.8 M$_\odot$ (triangles).
The figures at the left contain all the sample stars. The figures at the right
do not include the 9 bright stars of the cluster corona (see Sect. 3.2.1).}
\label{fig:6231cumulM}
\end{figure*}

The four diagrams of Fig. \ref{fig:6231cumulM} clearly indicate the existence
of mass segregation in NGC 6231. In the top two diagrams of Fig.
\ref{fig:6231cumulM} the mass intervals are set differently from those in the
bottom two panels. The two left hand diagrams of the same figure include the 9
bright stars of the cluster corona, while the right hand two diagrams do not
(see the caption of Fig. \ref{fig:6231cumulM} for interval limits).

Mass segregation is more pronounced for the massive stars (triangles), while
stars with masses in the range 5 $\leq$ $M$ $<$ 20 M$_\odot$ are
spatially well mixed (open squares and crosses). This latter population is however
more concentrated than the lower-mass population (filled squares).

From these curves, we conclude that only a dozen, bright, massive, mainly binary
stars are well concentrated toward the cluster center. The intermediate mass stars
(5 $<$ $M$ $<$ 20 M$_\odot$) are more uniformly distributed over the cluster area,
which means that mass segregation is not yet established over a rather large mass
interval.

\section{Discussion}

Our main goal, as stated in the Introduction, is to use these new results to test
the usual explanation of mass segregation in term of dynamical relaxation over a
large age interval. We then need to compare the radial structure of the three open
clusters (NGC 6231, Pleiades and Praesepe) and the observed mass segregation. We
shall also consider published results for a few other clusters (MonR2, Orion, M11,
M67).

\subsection{Relaxed clusters: Pleiades and Praesepe} 

Both clusters, respectively $10^{8}$ and 8 $\times$ 10$^{8}$ yr old, should be
well relaxed (the typical relaxation times for these open clusters are estimated
at around $10^{7}$ yr). As a consequence of equipartition of kinetic energy
between stars of different mass, both clusters should exhibit similar mass
segregation. We observe that this effect is alike for the most massive stars,
but appears less pronounced in Praesepe than in the Pleiades for the intermediate mass
stars, although M44 is about 8 times older than the Pleiades
(Fig. \ref{fig:cumulnorm3amas}).

Mathieu (1984) has examined the structure and mass segregation in NGC 6705 (M11)
on the basis of extensive proper motions and photometry. This cluster has an age
intermediate between that of the Pleiades and Praesepe ($\sim$ 2.3 $\times$ $10^{8}$
yr), with the mass of the most massive stars around 3.5  M$_{\odot}$. His Fig. 9
offers a clear evidence for a fine mass segregation and is very similar to our
Fig. \ref{fig:cumulnorm3amas}b (Pleiades).

\begin{figure}[thb]
\centerline{\psfig{figure=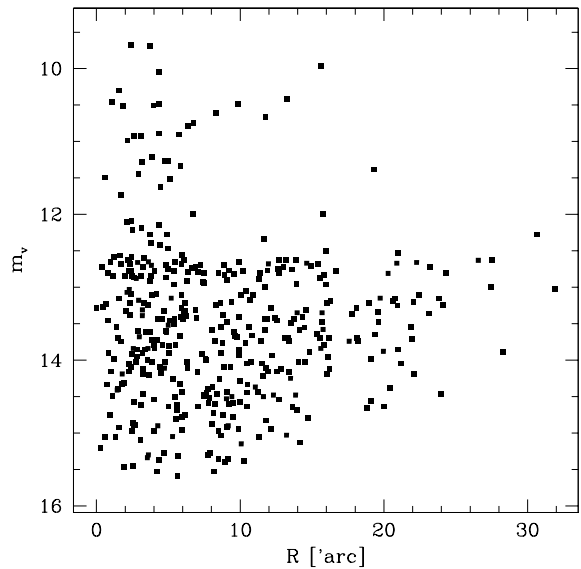,height=80mm,width=80mm}}
\caption[]{Apparent magnitudes of the stars as a function of their radial
distances to the center of M67. The HEP (Hydrogen Exhaustion Phase) gap is
apparent at $V \sim$ 13. This radial distribution of stars contrasts dramatically
with those presented for NGC 6231 and Praesepe (figs. \ref{fig:VvsR6231} and
\ref{fig:VvsR}).}
\label{fig:m67}
\end{figure}

The old open cluster M67 behaves quite differently: the radial distribution of the
member stars (Fig. \ref{fig:m67}) contrasts dramatically with those presented for
our three clusters. It presents a small amount of mass segregation for single stars
with $M$ $\sim$ 1.5 M$_{\odot}$. Only red giants, blue stragglers and binaries are
somewhat concentrated towards the cluster center (Mathieu 1985).
Figure \ref{fig:m67} also reveals the incompleteness of the membership list in the
outer part resulting either from the lower completeness of measurements of fainter
stars at large distance from the cluster center or from the membership estimates.

We will now consider two possible explanations for the less pronounced mass
segregation observed in Praesepe.

\subsubsection{The dominant mass component}

The first one follows the results of numerical simulations by Spitzer \& Shull
(1975). From them we infer that if stars belonging to a small range of
mass constitute almost all the cluster
mass, the spatial distribution of that component will be unaffected by interactions
with stars of other mass groups. Accordingly one should observe little, if any, mass
segregation among this dominant group. The other stars will be either more or
less concentrated, on whether they are heavier or lighter than the dominant group.
This kind of mass segregation will be only slightly dependant on the exact
individual stellar masses.

In the case of Praesepe, we note that the total cluster mass \textit{effectively}
observed (derived by summing up all the stellar masses) is contained within the
interval of the theoretically estimated masses (Table \ref{tab:masse}). It was not
the case for the Pleiades (RM98). We then observe a large part of the total mass
of Praesepe. If we consider that stars with masses between 0.9 and 2.3 M$_\odot$
constitute the dominant mass group of the cluster, we should observe little, or
no, mass segregation among this group. Furthermore, all heavier stars  ($M$  $>$
2.3 M$_\odot$) should be identically more concentrated.  This description would
explain correctly our results.

\subsubsection{The potential well}

The second explanation could be related to the smaller total mass of Praesepe
(Sect. 3.1.6), compared to that of the Pleiades (RM98). Praesepe has then
a shallower potential well than the Pleiades and the velocity distribution of the
stars of Praesepe are more severely truncated by the galactic tidal field (the two
clusters have similar galactic locations). This will result in a lesser degree of
mass segregation among Praesepe stars (Mathieu 1985).

However, we should keep in mind that the comparison between Praesepe and the
Pleiades has a limited validity, because the two clusters could have experienced
different external constraints. For instance, we could not exclude that the
lesser degree of mass segregation observed in Praesepe may be due to the effects
of external forces acting on the cluster.

\subsection{Non-relaxed cluster: NGC 6231}

The analysis of the structure of NGC 6231, the youngest open cluster that we
considered, clearly shows some mass segregation (Sect. 3.2.2). 

The estimation of the cluster relaxation time gives a value of about 10$^7$ yr
(Raboud 1997), larger than the cluster age (3-4 $\times$ 10$^6$ yr, Raboud et al.
1997). Therefore the cluster dynamical evolution did not have enough time to
produce energy equipartition among the cluster members and no mass segregation
should be present. Thus we are tempted to consider that the observed mass segregation
in NGC 6231 is initial and to identify it as a signature of the
stellar formation processes. Within this picture, the most massive stars form near
the cluster center.

However, as discussed in Raboud (1997), the computed relaxation is an upper limit.
This relaxation time, calculated with the standard equations from Chandrasekhar
(1942) and Spitzer \& Hart (1971), refers to stars of average mass. 
As real clusters present a wide mass spectrum, this implies that the systems
evolve on a timescale shorter than that estimated by this mean relaxation time. 
Furthemore, the relaxation time depends upon the location in the cluster: it
significantly increases from the center to the outer regions (Mathieu 1983).
Finally, $N$-body calculations that treat close gravitational encounters and
binary formation predict more rapid dynamical evolutions than that indicated by
the mean relaxation time (Sagar et al. 1988 and references therein).

We therefore cannot exclude a dynamical evolution on shorter timescales,
typically one order of magnitude, particularly in the innermost part of the
cluster or for the most massive stars.

Nevertheless, the mean relaxation time is also a lower limit because we observe only
the brightest stars of the cluster and therefore we underestimate the total number of
stars and the characteristic radius of the cluster while we overestimate its mean
stellar mass.

Numerical modelling are then truly needed to clearly quantify the amount of mass
segregation due to dynamical evolution and due to the initial conditions.

Such a modelling had been done by Bonnell \& Davies (1997) for the
Orion Nebula Cluster (ONC), based on the data of Hillenbrand (1997a). The authors show
that the position of massive stars in the center of rich young clusters cannot
be due to dynamical mass segregation. In particular, they claim that for producing
a Trapezium-like system within just a few crossing times, the massive stars most
likely formed within the inner 10\% of the cluster.

Other indications for an \textit{initial} mass segregation, i.e. an imprint of
the stellar formation processes and not a consequence of the cluster dynamical
evolution, have been obtained from the observations of other very young open
clusters like:
NGC 3293 (Herbst \& Miller 1982), NGC 6530 (McNamara \& Sekiguchi 1986), IC 1805
(Sagar et al. 1988), NGC 2264, NGC 6913, NGC 654, NGC 581, Tr 1 and h and $\chi$
Per (Pandey et al. 1991). But, as these indications of initial mass segregation are
mainly based on the comparison between the ages of the clusters and their mean
relaxation times, these studies suffer drawbacks similar to those described above.

Clusters still embedded within their parent molecular clouds and already displaying
mass segregation may be more convincing. Examples are, among others,
NGC 2024 and NGC 2071 (Lada \& Lada 1991). Such clusters have ages of
the order of their crossing time ($\sim 10^{6}$ yr) or below. Relaxation processes
are then negligible for them and the observed locations of their stars
are close to their birthplace. Consequently, the presence of mass segregation in these
\textit{extremely} young open clusters should not result from their dynamical
evolution.

All the preceding constatations favour the hypothesis that some of the mass
segregation observed in a cluster as young as NGC 6231 is likely to be primordial.

Inspection of Fig. \ref{fig:6231cumulM} also reveals that only the
\textit{most} massive stars are concentrated toward the cluster center. On the
contrary, stars with masses between $\sim$ 20 M$_\odot$ and $\sim$ 5 M$_\odot$
are spatially well mixed. Similar results are obtained for a cluster embedded
in the MonR2 molecular cloud (Carpenter et al. 1997). The authors
pointed out that mass segregation may be limited to the OB stars forming in this
region. Moreover, in the case of the ONC, Fig. 6 from Hillenbrand (1997b) shows
very different spatial distributions for stars more massive or less massive than
5 M$_\odot$. For masses smaller than 5 M$_\odot$ the distributions are rather
similar. We then conclude that, in very young clusters, mass segregation likely 
concerns only the most massive stars.

\subsubsection{Double origin for the mass segregation ?}

\begin{figure*}[thb]
\centerline{\psfig{figure=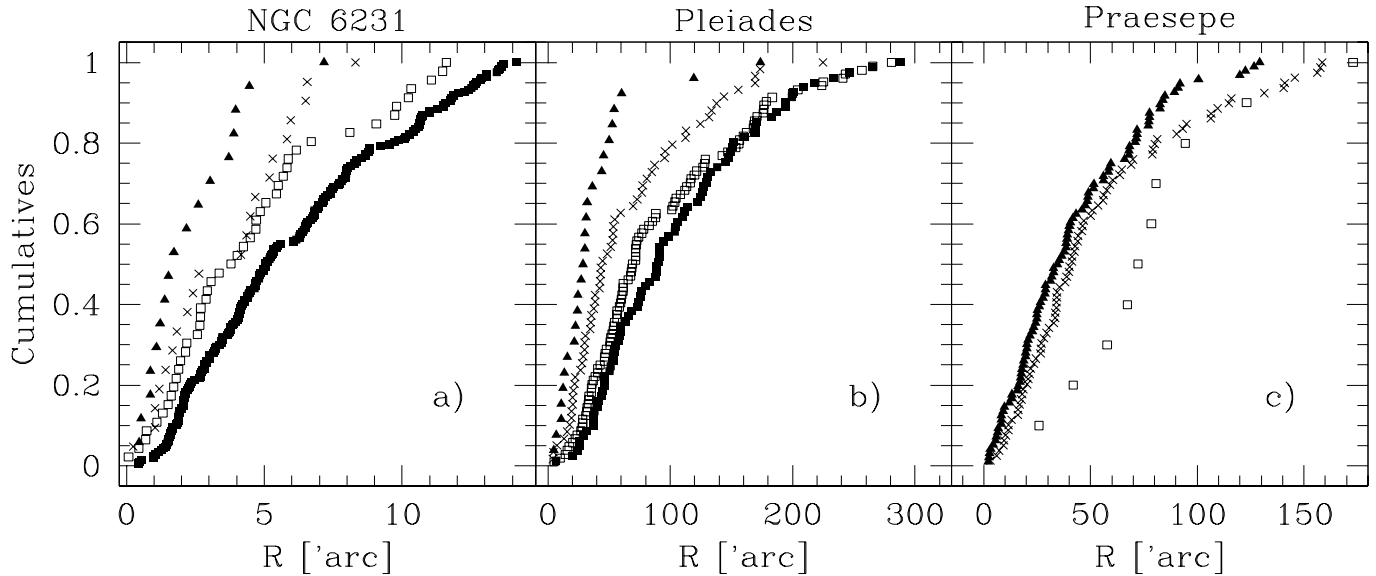,width=\textwidth}}
\caption[]{Cumulative distributions for stars in identical relative intervals of
mass, for the three clusters. These intervals are computed relatively to the maximum
stellar mass of the considered cluster. Triangles: $M$ $\geq$ 0.36 $\times$
M$_{max}$; crosses: 0.23 $\times$ M$_{max}$ $\leq$ $M$ $<$ 0.36 $\times$ M$_{max}$;
open squares: 0.14 $\times$ M$_{max}$ $\leq$ $M$ $<$ 0.23 $\times$ M$_{max}$; filled
squares: $M$ $<$ 0.14 $\times$ M$_{max}$. The 9 bright stars of the corona of
NGC 6231 are not included in the figure.}
\label{fig:cumulnorm3amas}
\end{figure*}

The evolutive picture emerging from the analysis of the considered
clusters (MonR2, Orion, NGC 6231, Pleiades, NGC 6705, Praesepe and M67) do not
agree with the usual description of the mass segregation, as a pure consequence of
dynamical evolution. We observe that the younger clusters (MonR2, Orion and
NGC 6231), likely still not relaxed, already present a mass segregation and that
the older ones (Praesepe, M67) present the lesser degree of mass segregation
(Fig. \ref{fig:cumulnorm3amas}). Possible explanations for the last observation
have been discussed in Sect. 4.1., but the presence
of some mass segregation within clusters likely still not relaxed implies a
reconsideration of the physical origin of this effect.

The above results allow us to propose a qualitative scenario for the
evolution of mass segregation with age in open clusters:

\textbf{(I)} The most massive stars ($M > 20$ M$_\odot$ for NGC 6231) form near the
center of clusters.

Several hypotheses could be made to explain the presence of massive stars near the
center of clusters at the early beginning of their life. Either the massive
protostars sink towards the center of clusters or physical conditions in the center
of protostellar clouds favour the formation of massive stars. These various hypothesis
are:  the dynamical friction between protostellar clouds and inter-protostellar
medium (Larson 1991, Gorti \& Bhatt 1995, 1996); the collision and coalescence of
protostellar clouds (Murray \& Lin 1996); the accretion of matter, during
stellar formation phases. This accretion could be faster in regions of higher
temperature and turbulence (Maeder 1997), i.e. in the center of protocluster
clouds, thus leading to the formation of more massive stars in these regions.
This last hypothesis implies that the IMF is dependent on the local physical
conditions. It is flatter in the central part of the cluster and steeper in the outer
part. Therefore open clusters could be the first physical environments, observed with
a sufficient spatial resolution, in which we note a non-universality of the IMF.

In the context of massive star formation in the center of clusters, it is worth
noting that we observe numerous examples of multiple systems of O-stars in the
center of very young open clusters. In the case of NGC 6231, 8 stars among the 10
brightest are spectroscopic binaries with periods shorter than 6 days. Moreover,
we observe trapezium systems of O-stars in the ONC, NGC 6823 and Tr 37. Four-component
and triple systems have also been found in NGC 2362 (van Leeuwen \& van Genderen 1997)
and Collinder 228 (Leung et al. 1979). 

\textbf{(II)} In less than $10^{7}$ yr these spatially concentrated massive stars
will disappear due to stellar evolution. As they may represent
a non-negligible percentage of the total mass of the cluster (between $\sim$10
and 30 \% in the case of NGC 6231), the disappearance of these massive
stars could lead to a violent relaxation phase. If a mass segregation was
previously established in the cluster it could be more or less erased during
this phase, depending on the importance of the initial population of massive
stars.

We are then {\it possibly} left with a cluster presenting {\it no} mass segregation
at all.
NGC 6531 (Forbes 1996) provides an example of such a cluster: it is
8 $\times$ 10$^{6}$ yr old and does not contain any stars with masses greater than
20 M$_\odot$ (which make up the concentrated population in NGC 6231). Forbes shows
convincingly that NGC 6531 does not exhibit any mass segregation, 
and he explains this result by the young age of the cluster. According
to him, NGC 6531 is too young for dynamical evolution to have left any significant 
impression. But this hypothesis was based on an estimation of the relaxation time
and suffers drawbacks described in the Sect. 4.2.

Another interesting point related to the disappearance of the massive stars
is the stability of the cluster. It is possible that a bound cluster
becomes unbound after this violent phase. Numerical simulations by Terlevich
(1987) show that clusters with flat initial mass functions have to be rich enough to
survive the initial violent period of mass loss.

\textbf{(III)} The last point of our scenario is that all  mass segregation
observed in older clusters (like the Pleiades or Praesepe) is \textit{merely} the
consequence of the cluster's \textit{dynamical evolution}. However, this conclusion
does not imply that NGC 6231 is a representative precursor of older clusters.

To better quantify this hypothesis of a possible double origin (initial
and dynamical) of the mass segregation we need to analyse the structure of open
clusters just old enough (around 10$^{7}$ yr) to have lost their most massive stars. 
Thus, one consequence of our hypothesis is that
some of these clusters, those which initially contained an important population  
of massive stars, should not present any mass segregation.

\section{Conclusion}

We present a study of the structure of Praesepe and NGC 6231. The results obtained,
compared with the Pleiades (RM98) and other clusters (Orion, Hillenbrand 1997a, b,
Bonnell \& Davies 1997; MonR2, Carpenter et al. 1997; M11, Mathieu 1984; M67,
Mathieu 1985) are used to discuss the mass segregation within open clusters. 

The study of the Praesepe structure has been performed on the basis
of the presently available data which limits the sample to stars brighter
than $V$ = 12. We used the best present knowledge on duplicity in Praesepe. Using
an asymmetry estimator applied to the apparent stellar positions we find that the
outer parts of Praesepe are round-shaped. This could be either real or only the
effect of projection. 

Praesepe is the second cluster (after the Pleiades in RM98) for which the mass
function of single stars and primaries of multiple systems have been determined
separately and compared. They turned out to be different.

We have observed mass segregation among cluster members (singles or multiples)
which does not depend on the binary periods. Consequently, binaries are more
concentrated than single stars and massive binaries are more concentrated than
less massive ones. However, the surprising result is that the mass segregation
observed in Praesepe (in the mass range 1.5-2.3 M$_\odot$) is less obvious than
in the Pleiades, although the former cluster is older.

Mass segregation is also observed in the very young open cluster NGC 6231. As
NGC 6231 is likely still not relaxed, this observation imply that the origin for
the mass segregation is possibly independant of the cluster dynamical evolution.
Moreover, we observe that only the most massive stars  ($M$  $>$ 20 M$_\odot$, in
NGC 6231) are centrally concentrated. The intermediate mass intervals are spatially
well mixed.

We therefore suggest that mass segregation observed in very young open clusters
concerns only the most massive stars and is mainly the signature of stellar
formation processes, implying a locally non-universal IMF, or of intra-cloud early
dynamical evolution. These massive stars disappear in less than 10$^{7}$ yr and this
phenomenon could lead to a violent relaxation phase, if the population of massive
stars is important. Then clusters with ages of the same order of magnitude could
present temporarily no mass segregation (like in NGC 6531, Forbes 1996), until
dynamical evolution becomes responsible for the settlement of mass segregation in
older, relaxed, clusters. In the oldest clusters, where the mass spectrum is much
narrower, only mass segregation between binaries and single stars should be
observable.

Therefore it appears important to underline that one cannot speak of mass
segregation in general, but one should indicate for each cluster which kinds of
stars are concentrated and which are not.

\end{document}